\documentclass[conference]{IEEEtran}
\usepackage{cite}

\ifCLASSINFOpdf
\usepackage[pdftex]{graphicx}
\else
\fi
\usepackage{amsmath}
\usepackage{amsfonts} 
\usepackage{amsthm}
\usepackage{amssymb}

\usepackage{BOONDOX-ds}

\usepackage{bbm}

\usepackage{kbordermatrix}

\newcommand{\pfun}{\mathop{\hbox{$\to$\kern-7pt\raise.9pt\hbox{\scalebox{1}[.55]{$|$}}\kern4pt} }}

\begin{document}

\title{Design, Generation, and Validation of \\ Extreme Scale Power-Law Graphs}

\author{\IEEEauthorblockN{
  Jeremy Kepner$^1$, Siddharth Samsi$^1$, 
  William Arcand$^1$, David Bestor$^1$, Bill Bergeron$^1$, Tim Davis$^2$, \\
  Vijay Gadepally$^1$, Michael Houle$^1$, Matthew Hubbell$^1$, Hayden Jananthan$^{1,3}$, Michael Jones$^1$, Anna Klein$^1$, \\
  Peter Michaleas$^1$, Roger Pearce$^4$, Lauren Milechin$^1$, Julie Mullen$^1$, Andrew Prout$^1$, \\
  Antonio Rosa$^1$, Geoff Sanders$^4$, Charles Yee$^1$, Albert Reuther$^1$
}
\vspace{1ex}
\IEEEauthorblockA{
	$^1$Massachusetts Institute of Technology,
    $^2$Texas A\&M,
    $^3$Vanderbilt University,
	$^4$Lawrence Livermore National Laboratory
}
}
\maketitle

\begin{abstract}
Massive power-law graphs drive many fields: metagenomics, brain mapping, Internet-of-things, cybersecurity, and sparse machine learning.  The development of novel algorithms and systems to process these data requires the design, generation, and validation of enormous graphs with exactly known properties.  Such graphs accelerate the proper testing of new algorithms and systems and are a prerequisite for success on real applications.  Many random graph generators currently exist that require realizing a graph in order to know its exact properties: number of vertices, number of edges, degree distribution, and number of triangles. Designing graphs using these random graph generators is a time-consuming trial-and-error process.  This paper presents a novel approach that uses Kronecker products to allow the exact computation of graph properties prior to graph generation.  In addition, when a real graph is desired, it can be generated quickly in memory on a parallel computer with no-interprocessor communication.  To test this approach, graphs with $10^{12}$ edges are generated on a 40,000+ core supercomputer in 1 second and exactly agree with those predicted by the theory.  In addition, to demonstrate the extensibility of this approach, decetta-scale graphs with up to $10^{30}$ edges are simulated in a few minutes on a laptop.
\end{abstract}

%
\IEEEpeerreviewmaketitle

\section{Introduction}
\let\thefootnote\relax\footnotetext{This material is based in part upon
work supported by the NSF under grant number DMS-1312831.  Any opinions,
findings, and conclusions or recommendations expressed in this material
are those of the authors and do not necessarily reflect the views of
the National Science Foundation.}

Power-law (or heavy-tail) \cite{Pareto1906,Zipf1935} graphs are found throughout a wide range of applications \cite{Barabasi1999,Gleich2015}.   In such graphs, there are a small number of vertices with a large number of edges and a large number of vertices with a small number of edges.  Specific domains where such graphs are important include genomics \cite{Morrison2005,Mooney2012,polychronopoulos2014conserved,Dodson2014,Dodson2015,gouda2016distribution}, brain mapping \cite{fornito2016graph}, computer networks \cite{BrinPage1998,Faloutsos1999,yan2015spectrum,fontugne2017scaling}, social media \cite{Zuckerburg2005,Kwak2009}, cybersecurity \cite{shao2015percolation,yu2015malware}, and sparse machine learning \cite{lee2008sparse,boureau2008sparse,glorot2011deep,yu2012exploiting,Kepner2017graphblasDNN}.

  Many graph processing systems are currently under development.  These systems are exploring innovations in algorithms \cite{Cormen2001,Miller2012,Buluc2014,voegele2017parallel,smith2017truss,hu2017trix,la2017ensemble,zhuzhunashvili2017preconditioned,low2017first,uppal2017scalable,mowlaei2017triangle}, software architecture \cite{BulucGilbert2011,Kepner2012-ch1,pearce2017triangle,halappanavar2017scalable,tom2017exploring,green2017quickly,kabir2017parallel,zhou2017design,hutchison2017distributed}, software standards \cite{Mattson2013,kepner2015graphs,kepner2016mathematical,bulucc2017design,davis2017graphblas}, and parallel computing hardware \cite{song2016novel,bisson2017static,date2017collaborative,debenedictis2017superstrider,manne2017if,kogge2017graph,gioiosa2017exploring}.
The development of novel algorithms and systems to process these data requires the design, generation, and validation of enormous graphs with known properties.  Such graphs accelerate the proper testing of new algorithms and systems and are a prerequisite for success on real applications.

Many random graph generators currently exist that require creating a graph in order to know its exact properties, such as the number of vertices, number of edges, degree distribution, and number of triangles.  Perhaps the most well-known and scalable power-law graph generator is used in the Graph500.org \cite{Chakrabarti2004,Leskovec2005,bader2006designing} and GraphChallenge.org \cite{dreher2016pagerank,ed,samsi2017static} benchmarks.  This generator, often referred to as R-MAT, is based on randomly sampling recursive Kronecker graphs.  Other highly scalable graph generators are based on randomly specified degree distributions \cite{seshadhri2012community,kepner2012perfect,gadepally2015using}.  Designing graphs using these random graph generators is an iterative process whereby the graph designer selects the parameters of the graph generator, randomly creates the graph with those parameters, and then measures the desired properties.  Such a process places certain natural limits on the ability of the graph designer to explore enormous graphs and know prior to graph generation the exact  properties of the graph.

This paper presents a complementary approach using Kronecker products that allows the exact computation of graph properties prior to graph generation.  In addition, when a real graph is desired, it can be generated quickly in memory on a parallel computer with no interprocessor communication. The paper begins  with a review of the relevant properties of Kronecker products.  Next, the types of constituent matrices that are suited for generating power-law graphs are described. Various mathematical properties of power-law Kronecker graphs are then derived.  Subsequently, a parallel algorithm for rapidly generating large graphs is provided.  A variety of performance results and specific examples of various graphs generated using this approach are presented.  Finally, the conclusions and a discussion of further research are given.

\section{Kronecker Products}

The Kronecker product of two square matrices is defined as follows \cite{VanLoan2000}
\begin{align*}
    &{\bf C} = {\bf A} \mathbin{\text{\textcircled{$\otimes$}}} {\bf B} = \\
     &\left(
     \begin{array}{cccc}
        {\bf A}(1,1) \otimes {\bf B} & {\bf A}(1,2) \otimes {\bf B} & ... & {\bf A}(1,m_A) \otimes {\bf B} \\
        {\bf A}(2,1) \otimes {\bf B} & {\bf A}(2,2) \otimes {\bf B} & ... & {\bf A}(2,m_A) \otimes {\bf B} \\
         \vdots   &  \vdots   & \ddots    & \vdots \\
        {\bf A}(m_A,1) \otimes {\bf B} & {\bf A}(m_A,2) \otimes {\bf B} & ... & {\bf A}(m_A,m_A) \otimes {\bf B} \\
     \end{array}  \right)
\end{align*}
where {\bf A}, {\bf B}, and {\bf C} matrices of scalar values ${\mathbb S}$
\begin{align*}
  {\bf A} & \in {\mathbb S}^{m_A \times m_A} \\
  {\bf B} & \in {\mathbb S}^{m_B \times m_B} \\
  {\bf C} & \in {\mathbb S}^{m_A m_B \times m_A m_B}
\end{align*}
More explicitly, the Kronecker product can be written as
  \[
    {\bf C}\bigl((i_A-1) m_A + i_B,(j_A-1) m_A + j_B\bigr) = {\bf A}(i_A,j_A) \otimes {\bf B}(i_B,j_B)
  \]
The element-wise multiply operation $\otimes$ can be a variety of functions so long as the resulting operation obeys the standard rules of element-wise multiplication, such as $0$ being the multiplicative annihilator for any value of $s \in {\mathbb S}$
\[
    0 \otimes s = s \otimes 0 = 0
\]
Furthermore, if element-wise multiplication and addition obey the conditions of a semiring \cite{gondran2007,golan2013,kepnerjananthan}, then the Kronecker product has many of the same desirable properties, such as associativity
  \[
    ({\bf A} \mathbin{\text{\textcircled{$\otimes$}}} {\bf B}) \mathbin{\text{\textcircled{$\otimes$}}} {\bf C} = 
    {\bf A} \mathbin{\text{\textcircled{$\otimes$}}} ({\bf B} \mathbin{\text{\textcircled{$\otimes$}}} {\bf C})
  \]
and element-wise distributivity over addition
 \[
    {\bf A} \mathbin{\text{\textcircled{$\otimes$}}} ({\bf B} \oplus {\bf C}) = 
    ({\bf A} \mathbin{\text{\textcircled{$\otimes$}}} {\bf B}) \oplus ({\bf A} \mathbin{\text{\textcircled{$\otimes$}}} {\bf C})
  \]
Finally, one unique feature of the Kronecker product is its relation to the matrix product. Specifically, the matrix product of two Kronecker products is equal to the Kronecker product of two matrix products
  \[
    ({\bf A} \mathbin{\text{\textcircled{$\otimes$}}} {\bf B}) ({\bf C} \mathbin{\text{\textcircled{$\otimes$}}} {\bf D})
     = 
     ({\bf A} {\bf C}) \mathbin{\text{\textcircled{$\otimes$}}} ({\bf B} {\bf D})
   \]
where matrix multiply 
  \[
  \mathbf{C} = \mathbf{A} \mathbf{B} = \mathbf{A} ~ {\oplus}.{\otimes} ~ \mathbf{B}
  \]
   is given by
  \[
   {\bf C}(i,j) = \bigoplus_k {\bf A}(i,k) \otimes {\bf B}(k,j)
  \]

\section{Generating Power-Law Graphs}
\begin{figure}[!t]
  \centering
    \includegraphics[width=3.5in]{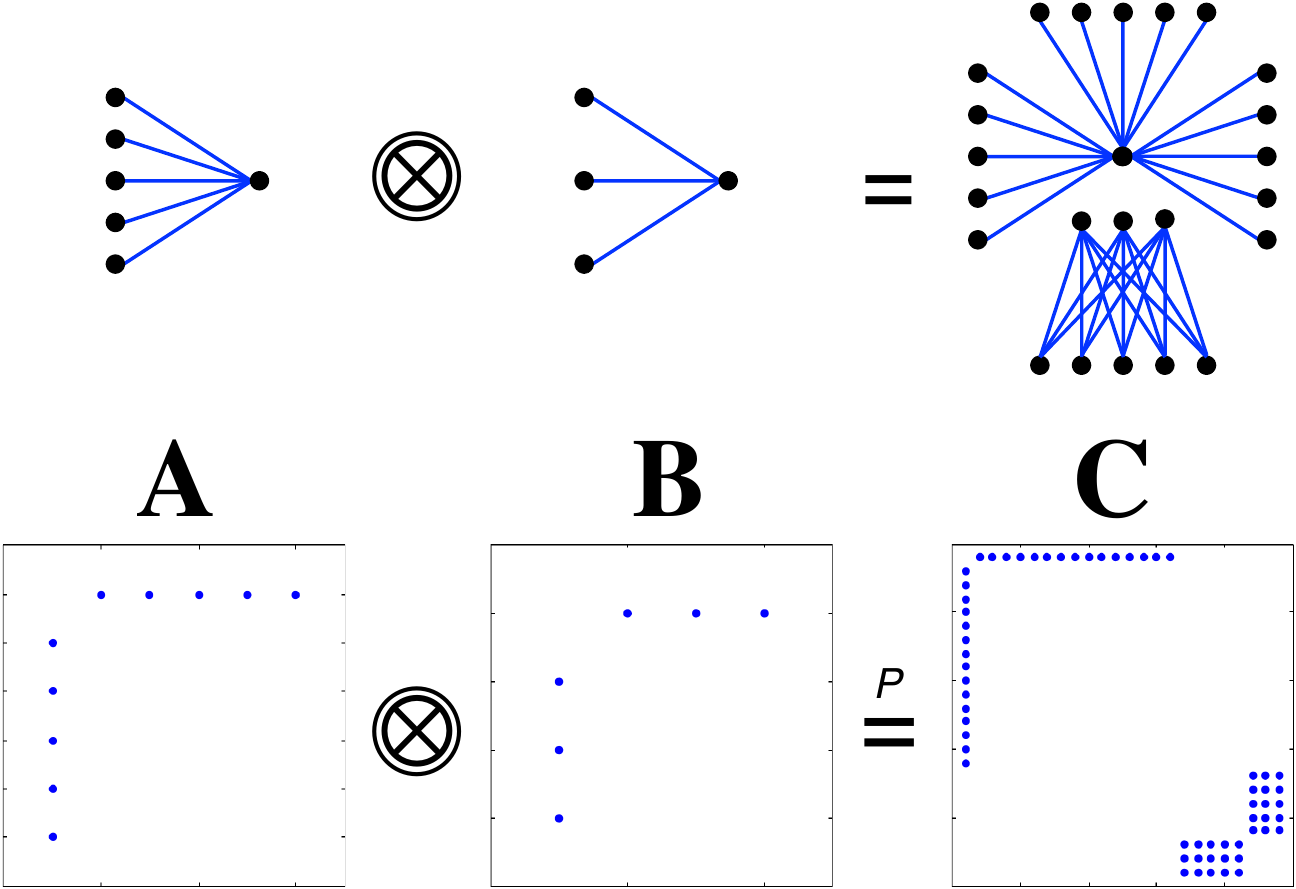}
       \caption{Kronecker product of the adjacency matrix of two bipartite graphs $\mathbf{A}$ and $\mathbf{B}$ results in a graph $\mathbf{C}$ with two bipartite sub-graphs.  The $\stackrel{P}{=}$ notation is used to indicate that the adjacency matrix $\mathbf{C}$ has been permuted so that the two bipartite sub-graphs are more apparent.}
      \label{fig:KroneckerProduct}
\end{figure}

Generating graphs is a common operation in a wide range of graph algorithms.  Graph generation is used in the testing of graph algorithms, in creating graph templates to match against, and for comparing real graph data with models.  Given a graph adjacency matrix $\mathbf{A}$, if
\[
  \mathbf{A}(i,j) = 1
\]
then there exists an edge going from vertex $i$ to vertex $j$ \cite{Konig1931,kepner2011graph}.  Likewise, if
\[
  \mathbf{A}(i,j) = 0
\]
then there is no edge from $i$ to $j$.  The Kronecker product of two graph adjacency matrices is a convenient, well-defined matrix operation that can be used for generating a wide range of graphs from a few parameters \cite{Chakrabarti2004,Leskovec2005}.  The relation of the Kronecker product to graphs is easily illustrated in the context of bipartite graphs.  Bipartite graphs have two sets of vertices, and every vertex has an edge to the other set of vertices but no edges within its own set of vertices.  The Kronecker product of such graphs was first looked at by Weischel \cite{Weischel1962}, who observed that the Kronecker product of two bipartite graphs resulted in a new graph consisting of two bipartite sub-graphs (see Figure~\ref{fig:KroneckerProduct}).

The essence of a power-law graph is that it has a degree distribution vector ${\bf n}(d)$ with non-zero entries that follows the relation
\[
     {\bf n}(d) \propto \frac{1}{d^\alpha}
\]
where $d$ is the number of edges in the vertex of a graph, ${\bf n}(d)$ is the number vertices with a specific degree $d$, and $\alpha > 0$ is the slope of the power law when it is plotted using logarithmic axes \cite{gadepally2015using}.  If the graph is represented as an adjacency matrix, then the degree of a vertex is the number of non-zero (nnz) entries in the corresponding row and column in the matrix.

A star graph is a bipartite graph where one set has only one vertex.  Star graphs are always a power-law graph. If a star graph has $m$ vertices, then the number of points in the star is given by
\[
    \hat{m} = m - 1
\]
with a corresponding degree distribution of
\begin{align*}
  {\bf n}(1) & = \hat{m} \\
  {\bf n}(\hat{m}) & = 1
\end{align*}
which agrees with the power-law relation $\alpha$ given by
\[
   \alpha = \frac{\log({\bf n}(1))}{\log(d_{\max})} = \frac{\log(\hat{m})}{\log(\hat{m})} = 1
\]
where $d_{\max}$ is the degree of the vertex with the most edges. 

The Kronecker product of two star graphs can, under certain conditions, produce another power-law graph.  In Figure~\ref{fig:KroneckerProduct}, the graph of the Kronecker product of two star graphs with $\hat{m}_A = 5$ and $\hat{m}_B = 3$ has a degree distribution of
\begin{align*}
  {\bf n}(1) & = 15 \\
  {\bf n}(3) & = 5 \\
  {\bf n}(5) & = 3 \\
  {\bf n}(15) & = 1
\end{align*}
which are all points on the curve
\[
     {\bf n}(d) = \frac{15}{d}
\]
The Kronecker product of star graphs can be used to build up extremely large power-law graphs.  The degree distributions will follow the power-law relation as long as all of the products of the corresponding $\hat{m}$ are unique.

  It is worth noting that real-world graphs often have approximate power-law distributions when plotted simply, as in this case, or when plotted with logarithmic degree binning, but rarely both.  It is possible to use Kronecker products to produce power-law graphs under logarithmic degree binning by placing additional constraints on the values of $\hat{m}$.

\section{Properties of Kronecker Graphs}
  The most powerful feature of Kronecker graphs is that many of their properties can be computed from their constituent matrices without ever having to form the full matrix.  It is thus possible to design and analyze extremely large graphs quickly and only actually form the full graph when it is needed.
 
  Let the adjacency matrix of graph ${\bf A}$ be constructed by the following Kronecker product
\[
     {\bf A} = {\Large \mathbin{\text{\textcircled{$\otimes$}}}}_{k=1}^{N_k} {\bf A}_k
\]
where ${\bf A}_k$ are each adjacency matrices of the smaller constituent graphs.  The number of vertices in the graphs is equal to the number of rows in ${\bf A}$ (or columns since ${\bf A}_k$ are square), which can be computed from
\[
    m_A = \prod_k m_{A_k}
\]
Likewise, the number of edges in the graph is equal to the number of non-zero entries in ${\bf A}$ and is given by
\[
    \text{nnz}({\bf A}) = \prod_k \text{nnz}({\bf A}_k)
\]
The degree distribution ${\bf n}_A(d)$ can be computed from the Kronecker product of the degree distributions ${\bf n}_{A_k}(d)$
\[
     {\bf n}_A(d) = {\Large \mathbin{\text{\textcircled{$\otimes$}}}}_{k=1}^{N_k} {\bf n}_{A_k}(d)
\]

\subsection{Triangles}
  The number of vertices, number of edges, and degree distribution are good examples of the core properties of Kronecker products.  A more sophisticated example is computing the number of triangles in a graph \cite{cohen2009,pavan2013,gilbert2015,burkhardt2016}.  Triangles are an important feature of a graph, and counting triangles is a basic property of many graph analysis systems.  The total number of triangles in a graph can be computed from the following formula
\[
N_\mathrm{tri}({\bf A}) = \frac{1}{6} \mathbbm{1}^{\sf T} ({\bf A} {\bf A} \otimes {\bf A}) \mathbbm{1}
\]
where $\mathbbm{1}$ is a column vector of all 1's and $\otimes$ is the element-wise product.  The same properties of Kronecker products apply to counting triangles, and the number of triangles can be computed from the component matrices via
\[
N_\mathrm{tri}({\bf A}) = \frac{1}{6} \prod_k \mathbbm{1}^{\sf T} ({\bf A}_k {\bf A}_k \otimes {\bf A}_k) \mathbbm{1}
\]

\begin{figure}[!t]
  \centering
    \includegraphics[width=3.5in]{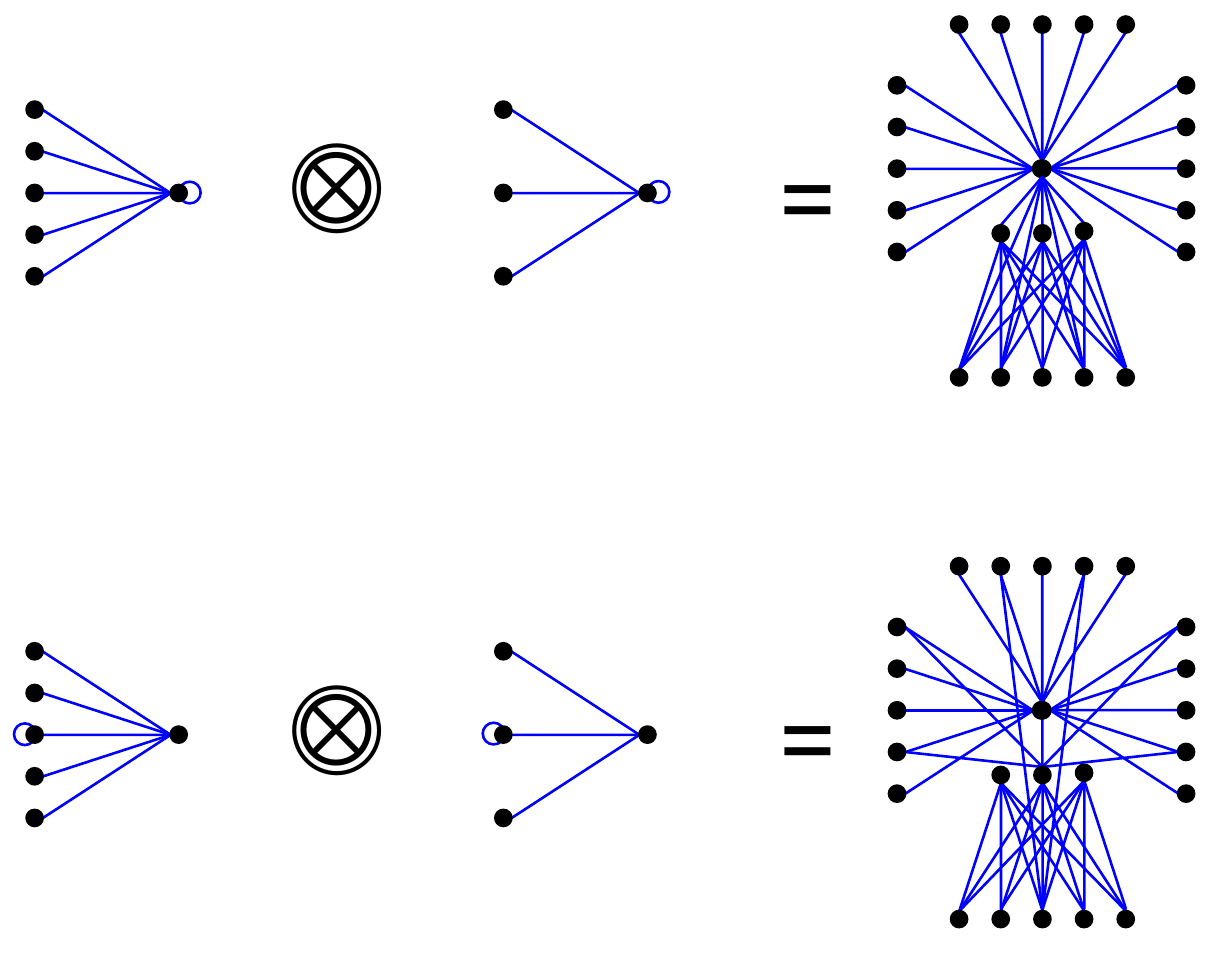}
       \caption{(top) Kronecker product of two star graphs with self-loops on the central vertex.  The resulting graph has 15 triangles. (bottom) Kronecker product of two star graphs with self-loops on a leaf node.  The resulting graph has 3 triangles.}
      \label{fig:KronCentralLeaf}
\end{figure}

\subsection{Case 1: Many Triangles}
  Bipartite graphs have no triangles, so the Kronecker product of star graphs will produce a large graph with zero triangles, which can be a useful test case.  Fortunately, it is possible to simply modify the ${\bf A}_k$ to create a graph with a rich triangle structure.  Specifically, if a self-loop is put on the central vertex of the star, the resulting graph will have a large number of triangles.  If the central vertex in the star is denoted by vertex 1, then a self-loop can be created in every constituent graph by setting
\[
    {\bf A}_k(1,1) = 1
\]
Removal of the self-loop in the final graph is accomplished by setting a single value back to zero
\[
    {\bf A}(1,1) = 0
\]
The number of vertices is unmodified by the inclusion of the self-loops.  The number of edges is computed from the ${\bf A}_k$ as before, followed by subtracting 1 from the total to account for the removal of the self-loop
\[
  \text{nnz}({\bf A}) - 1
\]
Likewise, the degree distribution is computed from the ${\bf A}_k$ as before with the following adjustments
\begin{align*}
  {\bf n}(d_{\max} -1) & = 1 \\
  {\bf n}(d_{\max}) & = 0
\end{align*}
The triangle count is computed from the ${\bf A}_k$ as before with the following correction
\[
    N_\mathrm{tri}({\bf A}) - \frac{1}{2} m_A + \frac{1}{3}
\]
Figure~\ref{fig:KronCentralLeaf} (top) shows an example of a graph with 15 triangles produced using this method.

\subsection{Case 2: Some Triangles}
A more modest number of triangles can be generated if one self-loop is put on one of the point vertices of each star, for example by setting
\[
    {\bf A}_k(m_{A_k},m_{A_k}) = 1
\]
Removal of the self-loop in the final graph is accomplished by setting a single value back to zero
\[
    {\bf A}(m_A,m_A) = 0
\]
The number of vertices is unmodified by the inclusion of the self-loops.  The number of edges is computed from the ${\bf A}_k$ as before, followed by subtracting 1 from the total to account for the removal of the self-loop
\[
   \text{nnz}({\bf A}) - 1
\]
Likewise, the degree distribution is computed from the ${\bf A}_k$ as before with the following adjustments
\begin{align*}
  {\bf n}(2^{N_k} -1) & = 1 \\
  {\bf n}(2^{N_k}) & = 0
\end{align*}
The triangle count is computed from the ${\bf A}_k$ as before with the following correction
\[
    N_\mathrm{tri}({\bf A}) - \frac{1}{2} 2^{N_k} + \frac{1}{3}
\]
Figure~\ref{fig:KronCentralLeaf} (bottom) shows an example of a graph with 1 triangle produced using this method.

\subsection{Incidence Matrix}
An incidence, or edge, matrix $\mathbf{E}$ uses the rows to represent every edge in the graph, and the columns represent every vertex.  There are a number of conventions for denoting an edge in an incidence matrix.  One such convention is to use two incidence matrices
\[
  \mathbf{E}_{\rm out}(e,i) = 1 \quad {\rm and} \quad \mathbf{E}_{\rm in}(e,j) = 1
\]
to indicate that edge $e$ is a connection from $i$ to $j$.  Incidence matrices are useful because they can easily represent multi-graphs and hyper-graphs.  These complex graphs are difficult to capture with an adjacency matrix.  One of the most common uses of matrix multiplication is to construct an adjacency matrix from an incidence matrix representation of a graph.  For a graph with out-vertex incidence matrix $\mathbf{E}_\mathrm{out}$ and in-vertex incidence matrix $\mathbf{E}_\mathrm{in}$, the corresponding adjacency matrix is \cite{jananthan2017constructing,kepnerjananthan}
\[
  \mathbf{A} = \mathbf{E}_\mathrm{out}^{\sf T} \mathbf{E}_\mathrm{in}
\]
Kronecker products can also be used to construct incidence matrices that satisfy the above adjacency matrix equation.  Specifically, let $\mathbf{E}_{k,\mathrm{out}}$ and $\mathbf{E}_{k,\mathrm{in}}$ be incidence matrices corresponding to $\mathbf{A}_k$.  The incidence matrices can then be constructed by 
\[
     {\bf \mathbf{E}_\mathrm{out}} = {\Large \mathbin{\text{\textcircled{$\otimes$}}}}_{k=1}^{N_k} \mathbf{E}_{k,\mathrm{out}}
\]
and
\[
     {\bf \mathbf{E}_\mathrm{in}} = {\Large \mathbin{\text{\textcircled{$\otimes$}}}}_{k=1}^{N_k} \mathbf{E}_{k,\mathrm{in}}
\]
It is worth noting that the order of edges in the incidence matrices is not uniquely determined. Different realizations of an incidence matrix are only equivalent when comparing their resulting adjacency matrices.

\section{Parallel Generation}
  Kronecker products allow the properties of a graph to be determined in advance, thus avoiding the iterative approach of other methods.  Once the desired graph properties have been determined, Kronecker products also allow large graphs to be generated quickly on a parallel processor.  The overall approach is to split the constituent matrices into two matrices ${\bf B}$ and ${\bf C}$ 
\begin{align*}
     {\bf A} & = {\Large \mathbin{\text{\textcircled{$\otimes$}}}}_{k=1}^{N_k} {\bf A}_k \\
             & = \left({\Large \mathbin{\text{\textcircled{$\otimes$}}}}_{k=1}^{N_B} {\bf A}_k \right)
               \mathbin{\text{\textcircled{$\otimes$}}}
               \left({\Large \mathbin{\text{\textcircled{$\otimes$}}}}_{k=N_B - 1}^{N_C} {\bf A}_k \right) \\
             & = {\bf B} \mathbin{\text{\textcircled{$\otimes$}}} {\bf C}
\end{align*}
The matrices ${\bf B}$ and ${\bf C}$ are designed so that both can fit in the memory of any one processor.  Let the parallel computer have $N_p$ processors, and each processor is given an identifier $p$ \cite{Bliss2006,Kepner2009}.  Each processor reads in ${\bf B}$ and ${\bf C}$ and extracts the triples of the non-zero element ${\bf B}$ into three vectors ${\bf i}$, ${\bf j}$, and ${\bf s}$, each of length $\textrm{nnz}({\bf B})$.  Each processor then selects a $\mathrm{nnz}({\bf B})/N_p$  of the triples ${\bf i}_p$, ${\bf j}_p$, and ${\bf s}_p$.  If the underlying sparse storage of the matrices is compressed sparse columns (CSC), then the minimum value of ${\bf j}_p$ is subtracted from ${\bf j}_p$ and a new matrix ${\bf B}_p$ is formed from these triples.  Each processor can then form the submatrix ${\bf A}_p$ of the overall matrix ${\bf A}$ via the Kronecker product
\[
  {\bf A}_p = {\bf B}_p \mathbin{\text{\textcircled{$\otimes$}}} {\bf C}
\]
The resulting ${\bf A}_p$ matrices will have the same number of non-zero entries on each processor.   In addition, the resulting graph is free of many of the problematic vertices and edges, such as empty vertices and self-loops, that are found in randomly generated graphs.  These problematic vertices and edges often require  randomly generated graphs to be reindexed before their properties can be computed.

\section{Results}

\begin{figure}[!t]
  \centering
    \includegraphics[width=3.5in]{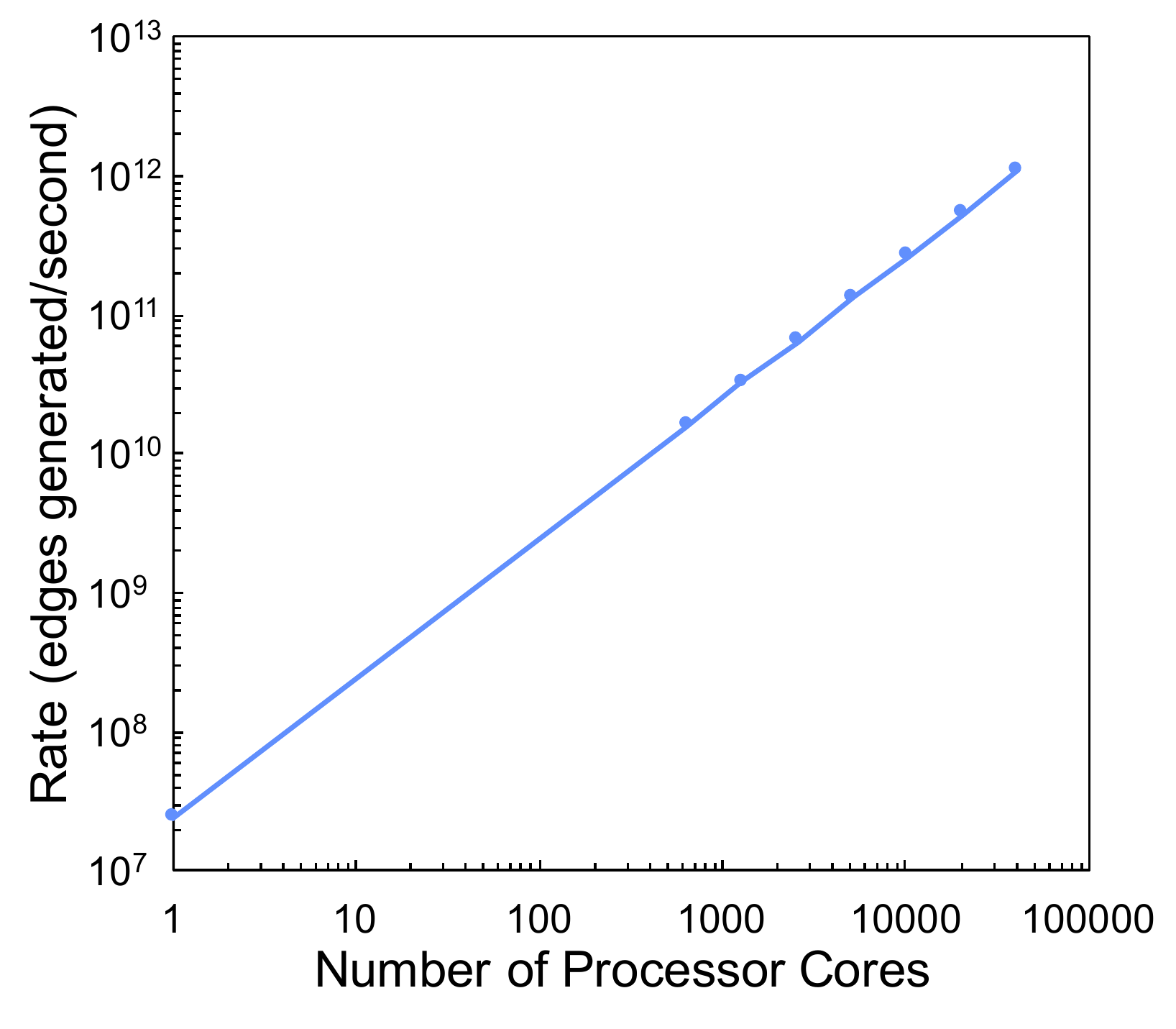}
       \caption{Edge generation rate vs. number of processor cores.  Performance scales linearly with processor cores and achieves a peak rate of over 1 trillion edges generated per second on over 40,000 processor cores.}
      \label{fig:Speedup}
\end{figure}

  This section presents a variety of scalability results to demonstrate the properties of the proposed Kronecker graph generation method.  Figure~\ref{fig:Speedup} shows the rate of graph edge generation as a function of the number of processing cores used in the parallel graph generation technique described in the previous section.  In this example, ${\bf B}$ is a 530,400 vertex graph with 13,824,000 edges constructed from the Kronecker product of star graphs with $\hat{m} = \{3, 4, 5, 9, 16\}$.  Likewise, ${\bf C}$ is a 21,074 vertex graph with 82,944 edges constructed from the Kronecker product of star graphs with $\hat{m} = \{81, 256\}$.  The Kronecker product of ${\bf B}$ and ${\bf C}$, produces a graph ${\bf A}$ with 11,177,649,600 vertices and 1,146,617,856,000 edges and zero triangles.  This graph construction was run in parallel on a supercomputer consisting of 648 compute nodes, each with at least 64 Xeon processing cores, for a total of 41,472 processing cores.  Using the entire system, the trillion edge graph was generated in 1 second.

\begin{figure}[!t]
  \centering
    \includegraphics[width=3.5in]{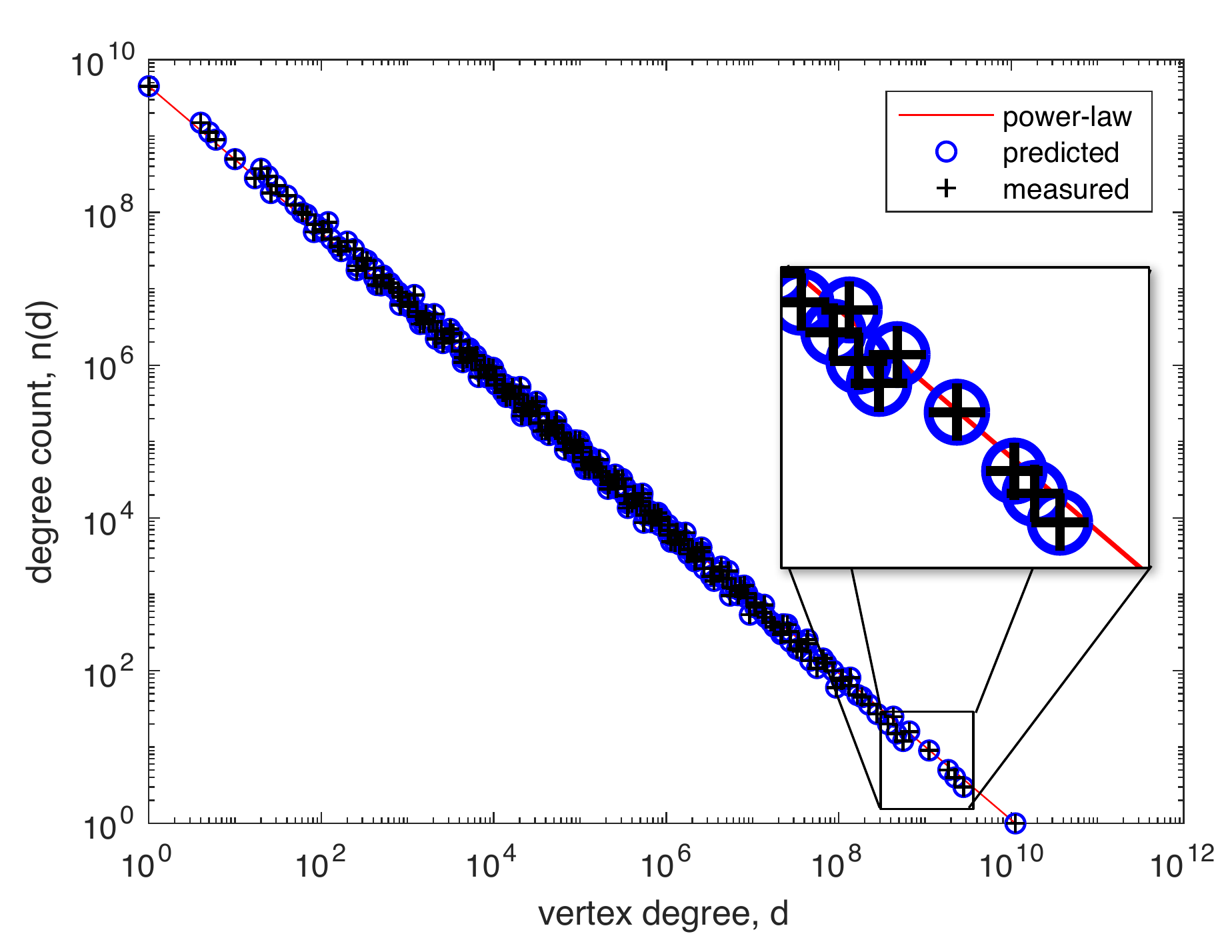}
       \caption{Trillion-edge ($10^{12}$) power-law Kronecker graph showing the exact agreement between the predicted and measured degree distribution.  The resulting graph has exactly 11,177,649,600 vertices, 1,853,002,140,758 edges, and 6,777,007,252,427 triangles.}
      \label{fig:TrillionEdgeVerified}
\end{figure}

  Computing the degree distribution of the generated graph can be used to verify that a generated graph agrees with the theory.  Figure~\ref{fig:TrillionEdgeVerified} shows the measured and predicted degree distribution of a graph produced using the parallel graph generation technique.  In this example, ${\bf B}$ is a 530,400 vertex graph with 22,160,060 edges constructed from the Kronecker product of star graphs with $\hat{m} =\{3, 4, 5, 9, 16\}$ and self-loops on the central vertices of the stars.  Likewise, ${\bf C}$ is a 21,074 vertex graph with 83,618 edges constructed from the Kronecker product of star graphs with $\hat{m} = \{81, 256\}$ and self-loops on the central vertices of the stars.  The Kronecker product of ${\bf B}$ and ${\bf C}$ produces a graph ${\bf A}$ with 11,177,649,600 vertices and 1,853,002,140,758 edges and 6,777,007,252,427 triangles.  This calculation confirms that the predicted and measured graph are in exact agreement.

\begin{figure}[!t]
  \centering
    \includegraphics[width=3.5in]{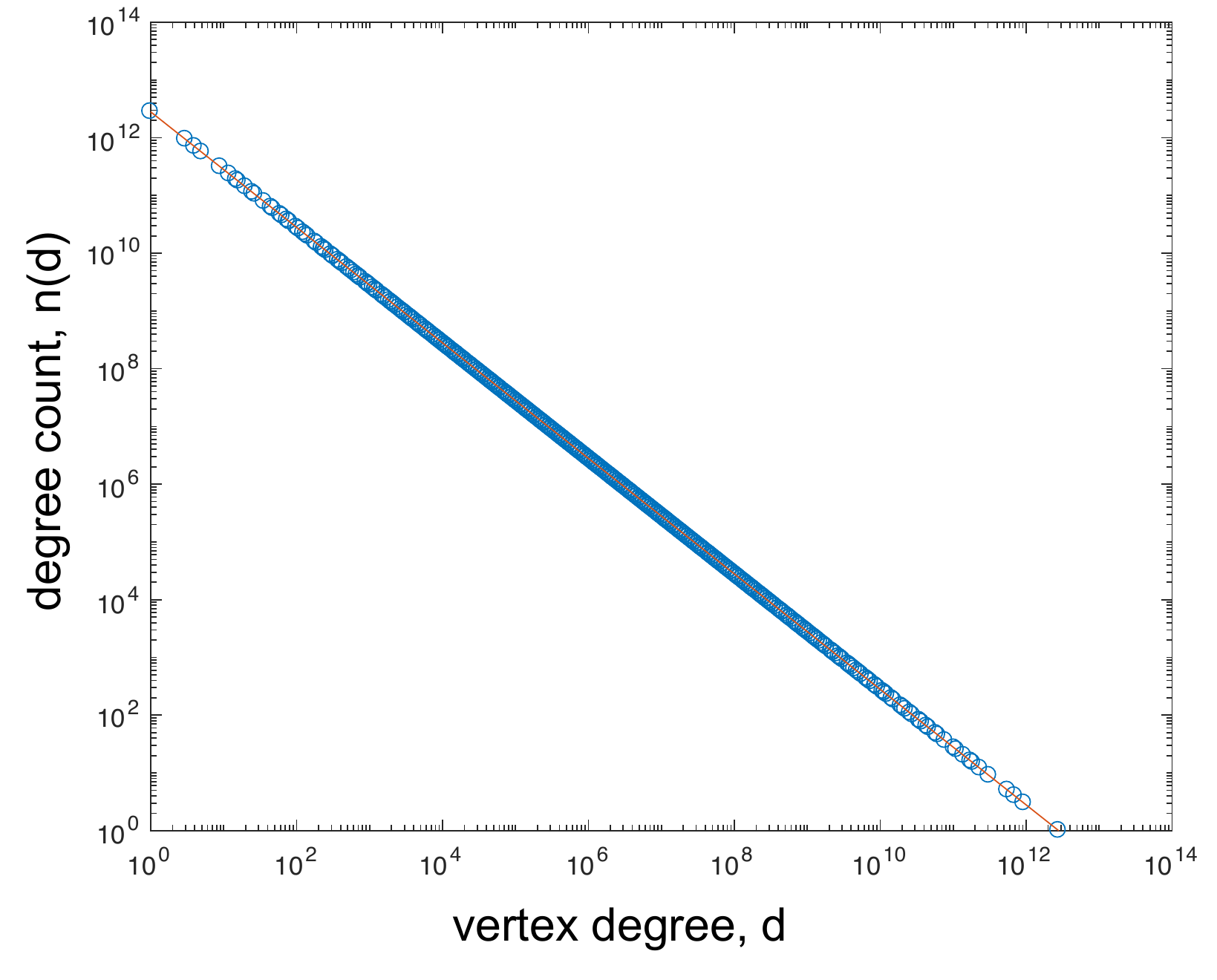}
       \caption{Quadrillion-edge ($10^{15}$) power-law Kronecker graph predicted degree distribution.  The resulting graph has exactly 6,997,208,649,600 vertices, 1,433,272,320,000,000 edges, and zero triangles.}
      \label{fig:NoTriDegreeDistribution}
\end{figure}

\begin{figure}[!t]
  \centering
    \includegraphics[width=3.5in]{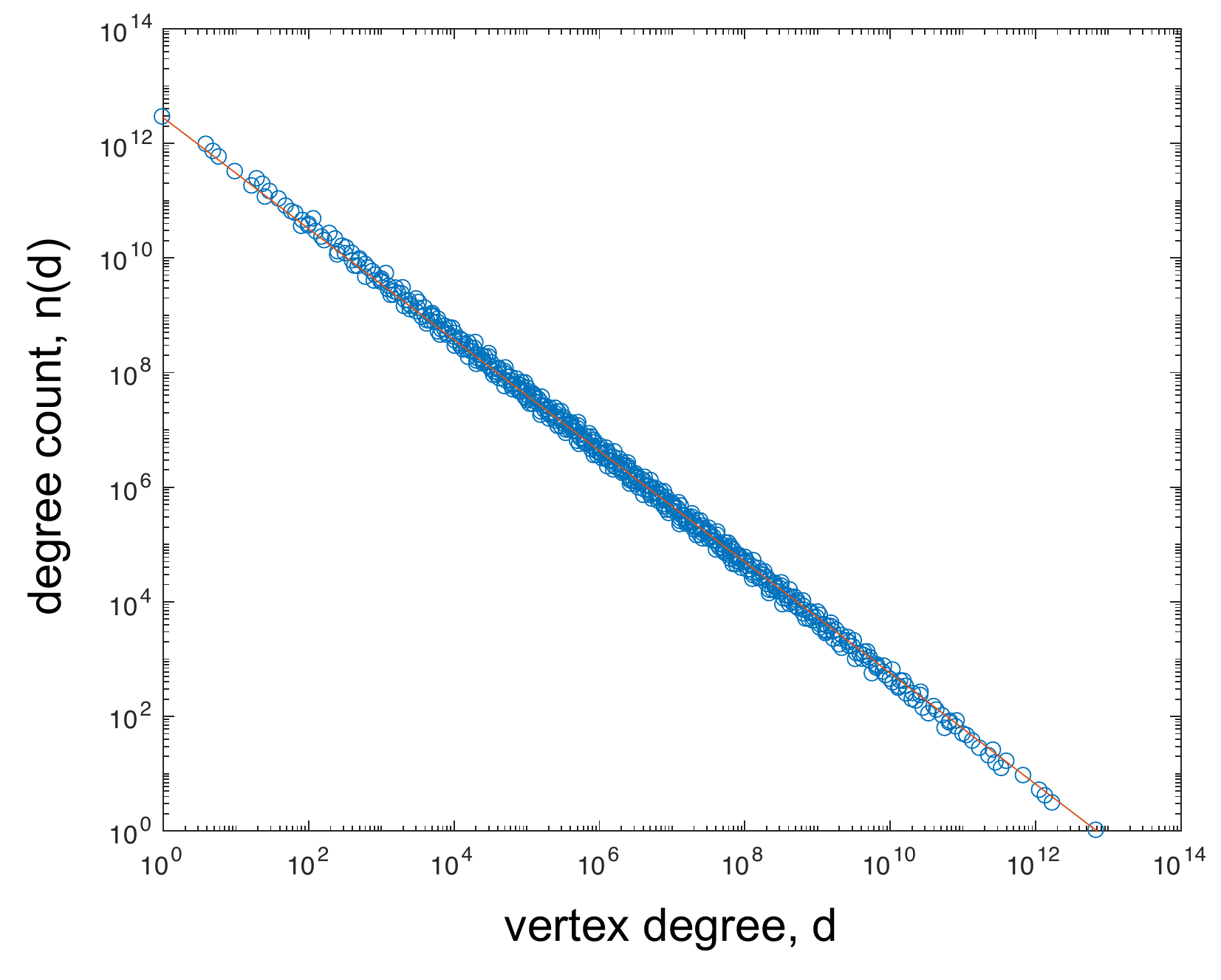}
       \caption{Quadrillion-edge ($10^{15}$) power-law Kronecker graph predicted degree distribution.  The resulting graph has exactly 6,997,208,649,600 vertices, 2,318,105,678,089,508 edges, and 12,720,651,636,552,426 triangles.}
      \label{fig:ManyTriDegreeDistribution}
\end{figure}

  Kronecker products can allow the rapid design of very large graphs suitable for the world's largest computers.  Figures~\ref{fig:NoTriDegreeDistribution} and \ref{fig:ManyTriDegreeDistribution} show the degree distribution for two graphs with over $10^{15}$ edges.  Both graphs are generated from star graphs with $\hat{m} =\{3, 4, 5, 9, 16, 25, 81, 256, 625\}$ and 6,997,208,649,600 vertices.  Figure~\ref{fig:NoTriDegreeDistribution} has 1,433,272,320,000,000 edges and zero triangles, and the degree distribution exactly follows the power-law degree formula.  Figure~\ref{fig:ManyTriDegreeDistribution} is generated with self-loops on the central vertices producing 2,318,105,678,089,508 edges, 12,720,651,636,552,426 triangles, with the degree distribution that follows the power-law degree formula with small deviations above and below the line.

\begin{figure}[!t]
  \centering
    \includegraphics[width=3.5in]{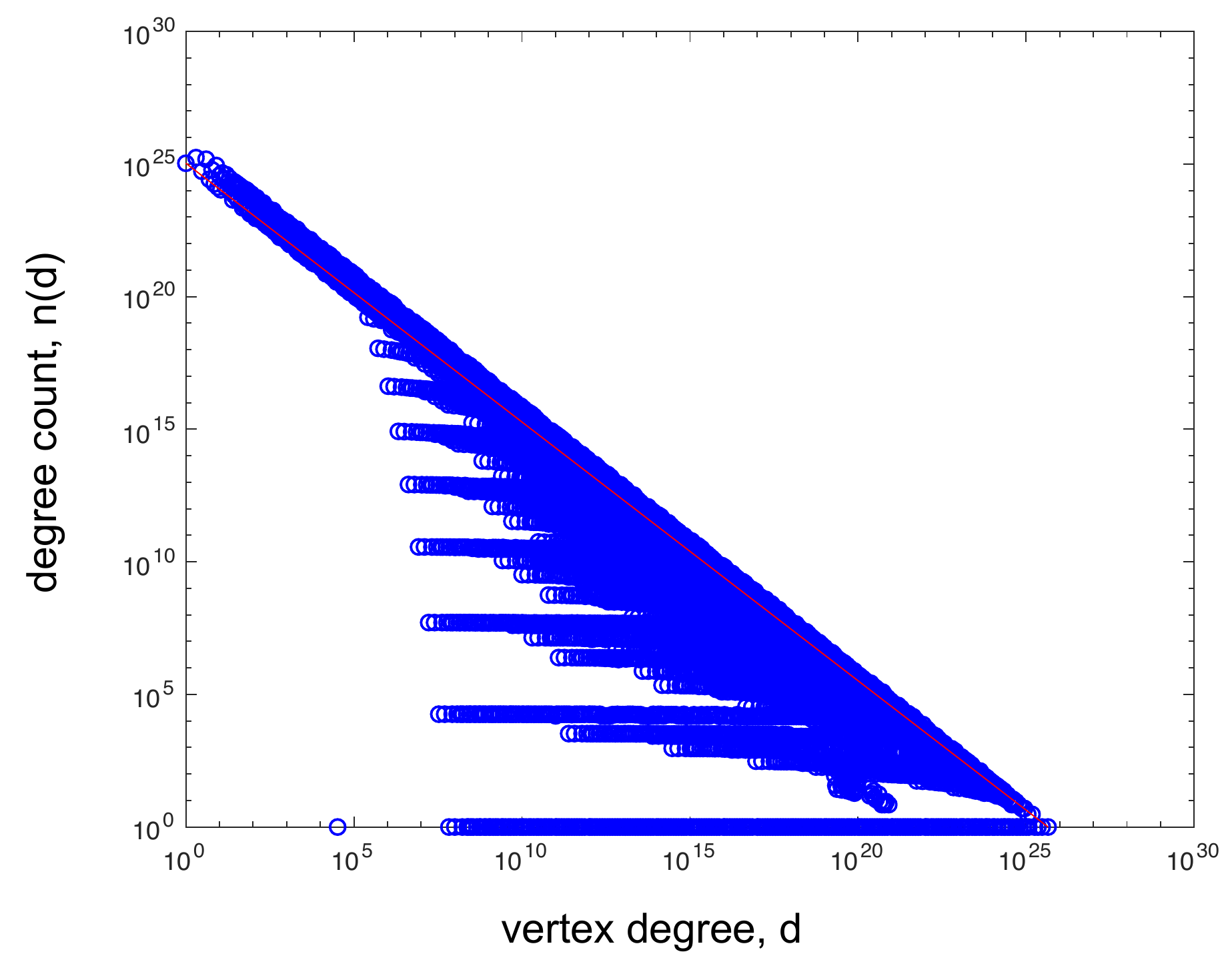}
       \caption{Predicted degree distribution of a decetta-edge ($10^{30}$) power-law Kronecker graph.  The resulting graph is predicted to have exactly 144,111,718,793,178,936,483,840,000 vertices,  2,705,963,586,782,877,716,483,871,216,764 edges, and 178,940,587 triangles.}
      \label{fig:SomeTriDegreeDistribution}
\end{figure}

  Kronecker products can also enable the exact analysis of graphs that are far beyond the scale of any current or planned computing system.  Figures~\ref{fig:SomeTriDegreeDistribution} shows the degree distribution of a graph with over $10^{30}$ edges.  The graph was generated from star graphs with $\hat{m} =\{3, 4, 5, 7, 11, 9, 16, 25, 49, 81, 121, 256, 625, 2401, 14641\}$ and a self-loop on one point vertex of each star.  The resulting graph has exactly 144,111,718,793,178,936,483,840,000 vertices, 2,705,963,586,782,877,716,483,871,216,764 edges, and 178,940,587 triangles.  Most of the points follow the power-law degree line, but there are many points that deviate from this.  This degree distribution was computed on a standard laptop computer in a few minutes.

\section{Conclusion}

Emerging data in metagenomics, brain mapping, Internet-of-things, cybersecurity, and sparse machine learning produce massive power-law graphs and are driving the development of novel algorithms and systems to process these data.  The scale and distribution of these data makes validation of graph processing systems a significant challenge.  The ability to create enormous graphs with exactly known properties can significantly accelerate the design, generation, and validation of new graph processing systems.  Many current graph generators produce random graphs whose exact properties, such as number of vertices, number of edges, degree distribution, and number of triangles, can only be computed after the graph has been generated.  Thus, designing graphs using these random graph generators is a time-consuming trial-and-error process.

Kronecker products of the adjacency matrices of star graphs are a powerful way to create large power-law graphs.  The properties of Kronecker products allow many properties of a larger graph to be computed by simply combining the corresponding properties of the constituent matrices.  The ability to compute the properties of large graphs using only small graphs allows the graph designer to find these prior to creating the actual graph.  Furthermore, real graphs can be created using Kronecker products on a parallel computer with no interprocessor communication. The resulting graphs will have the same number of edges on each processor.   In addition, the graph avoids many of the difficulties, such as empty vertices and self-loops, that are found in other graph generators that rely random sampling.  These problematic vertices and edges often require randomly generated graphs to be reindexed before their properties can be computed.

To test this approach, graphs with $10^{12}$ edges are generated on a 40,000+ core supercomputer in 1 second and exactly agree with those predicted by the theory.  In addition, in order to demonstrate the extensibility of this approach, decetta-scale graphs with up to $10^{30}$ edges are simulated in a few minutes on laptop.  These results indicate that the proposed method can be a powerful tool for enabling the design, generation, and validation of new graph processing systems.

This paper has presented formulas for a number of properties of Kronecker graphs.  There are many additional properties that could be computed in future research, such as eigenvectors, iso-parametric ratios, betweenness centrality, and triangle enumeration.  The parallel Kronecker graph generator is ideally suited to the GraphBLAS.org software standard and the creation of a high performance version using this standard is a future goal.  Finally, the ability to reason about graphs that are beyond any current or planned computer opens up new possibilities for the theoretical study of phenomena on these large graphs.


\section*{Acknowledgments}
%
%

The authors wish to acknowledge the following individuals for their contributions and support: Alan Edelman, Charles Leiserson, Steve Pritchard, Michael Wright, Bob Bond, Dave Martinez, Sterling Foster, Paul Burkhardt, and Victor Roytburd.



\bibliographystyle{IEEEtran}
\bibliography{aarabib}
%

\end{document}